\documentclass[aps,prx,twocolumn,superscriptaddress]{revtex4-1}
\usepackage{amsmath}
\usepackage{amssymb}
\usepackage{color}
\usepackage{graphicx}
\usepackage{epstopdf}

\DeclareMathOperator{\arsinh}{arsinh}
\DeclareMathOperator{\artanh}{artanh}

\begin{document}

%\title{Achieving large zeta-potentials with electrolyte-infused porous surfaces}
\title{Achieving large zeta-potentials with charged porous surfaces}

\author{Olga I. Vinogradova}
\email[Corresponding author: ]{oivinograd@yahoo.com}

\affiliation{Frumkin Institute of Physical Chemistry and Electrochemistry, Russian Academy of Sciences, 31 Leninsky Prospect, 119071 Moscow, Russia}
\affiliation{DWI - Leibniz Institute for Interactive Materials,  Forckenbeckstr. 50, 52056 Aachen, Germany}

\author{Elena F. Silkina}
\affiliation{Frumkin Institute of Physical Chemistry and Electrochemistry, Russian Academy of Sciences, 31 Leninsky Prospect, 119071 Moscow, Russia}

\author{Naren Bag}
\affiliation{DWI - Leibniz Institute for Interactive Materials,  Forckenbeckstr. 50, 52056 Aachen, Germany}

\author{Evgeny S. Asmolov}
\affiliation{Frumkin Institute of Physical Chemistry and Electrochemistry, Russian Academy of Sciences, 31 Leninsky Prospect, 119071 Moscow, Russia}

\date{\today }

\begin{abstract}

We discuss an electro-osmotic flow near  charged porous coatings of a finite hydrodynamic permeability, impregnated with an outer electrolyte solution. It is shown that their electrokinetic (zeta) potential is generally augmented compared to the surface electrostatic potential, thanks to a large liquid slip at their surface emerging due to an electro-osmotic flow in the enriched by counter-ions porous films. The inner flow
shows a very rich behavior controlled by the volume charge density of the coating, its Brinkman length, and concentration of added salt. Interestingly, even for relatively small Brinkman length the zeta-potential can, in some cases, become huge, providing a very fast outer flow in the bulk electrolyte. When the Brinkman length is large enough, the zeta-potential could be extremely high even at practically vanishing surface potential.
   To describe the slip velocity in a simple manner, we introduce a concept of an electro-osmotic slip length and demonstrate that the latter is always defined by the hydrodynamic permeability of the porous film, and also, depending on the regime, either by its volume charge density or
   the salt concentration. These results provide a framework for the rational design of porous coatings to enhance electrokineic phenomena,
and for tuning their properties by adjusting bulk electrolyte concentrations, with direct applications in
microfluidics.
\end{abstract}

\maketitle

\renewcommand{\vec}[1]{\boldsymbol{#1}}

\affiliation{DWI - Leibniz Institute for Interactive Materials,
Forckenbeckstr. 50, 52056 Aachen, Germany}

\affiliation{Frumkin Institute of Physical Chemistry and Electrochemistry,
Russian Academy of Sciences, 31 Leninsky Prospect, 119071 Moscow, Russia}

\affiliation{Frumkin Institute of Physical Chemistry and Electrochemistry,
Russian Academy of Sciences, 31 Leninsky Prospect, 119071 Moscow, Russia}
\affiliation{Lomonosov Moscow State University, 119991 Moscow, Russia}

\affiliation{DWI - Leibniz Institute for Interactive Materials,
Forckenbeckstr. 50, 52056 Aachen, Germany}
\affiliation{Frumkin Institute of
Physical Chemistry and Electrochemistry, Russian Academy of Sciences, 31
Leninsky Prospect, 119071 Moscow, Russia}
\affiliation{Lomonosov Moscow State University, 119991 Moscow, Russia}

%\keywords{one, two, three, four}
%\begin{keyword}
%Charged soft particle, Dielectric core, Electric polarization, Mobility reversal, Numerical computation.
%\end{keyword}
%\end{frontmatter}

%\section{Introduction}

\section{Introduction}\label{sec:introduction}

When an  electric field $E$ is applied tangent to a charged surface, an electro-osmotic flow of an electrolyte solution is induced~\cite{anderson.jl:1989}. The successful understanding of electro-osmosis, due to Smoluchowski~\cite{smoluchowski.m:1921}, was a triumph of 20th century colloid physics. Smoluchowski clarified that the electro-osmotic flow takes its  origin in the  adjacent diffuse layer of counter-ions, and argued that the finite $V_{\infty}$ outside of the diffuse layer is given by~\cite{smoluchowski.m:1921}
\begin{equation}\label{eq:smoluchowsky}
  V_{\infty} = - \dfrac{\varepsilon E}{4 \pi \eta } Z,
\end{equation}
with permittivity of the solution $\varepsilon$, its dynamic
viscosity $\eta$, and so-called (electro-hydrodynamic) zeta-potential  $Z$ of the surface, where the no-slip boundary condition is postulated. This postulate implies, via the Stokes equation, that $Z$ must be equal to the surface (electrostatic) potential $\Psi_s$.

This classical subject of colloid and interface science is currently widely used in a microfluidics that requires manipulating fluids in thin channels~\cite{squires.tm:2005}. However, the no-slip surfaces and a consequent concept of $Z$ as $\Psi_s$  that form the basis of the classical theory are of limited applicability.
Since typical values of $\Psi_s$ are of the order of a few tens of mV, to achieve
velocities of a few millimeters per second a high-voltage supply  is required, which is an obvious impediment for the use~\cite{stone.ha:2004}. Consequently, a search for mechanisms for  generating a large zeta-potential in the low-voltage situation is one of the important  challenges in the modern microfluidics. A natural and promising strategy would be a generation of a finite liquid velocity at the surface that may help to augment $Z$ to a very large value. An increase in $V_{\infty}$ can be then quantified by $\mathcal{A} =  Z/\Psi_s$, we refer below to as an amplification factor.

One avenue to increase $Z$ and $\mathcal{A}$ is to employ a hydrodynamic (hydrophobic) slippage~\cite{vinogradova.oi:1999}. This is usually quantified by the hydrodynamic slip length $b$, which can be of the order of tens of nanometers~\cite{charlaix.e:2005,vinogradova.oi:2003,joly.l:2006,vinogradova.oi:2009}, but not much more. Simple arguments show that $V_{\infty}$ can be amplified by a factor of~\cite{silkina.ef:2019,muller.vm:1986,joly.l:2004}
\begin{equation}\label{eq:hydrophobic}
 \mathcal{A} = 1  + \dfrac{2 \kappa b}{\psi_{s} } \sinh \left(\dfrac{\psi_{s}}{2}\right),
\end{equation}
where dimensionless $\psi_{s} = \textsl{e} \Psi_s/(k_{B}T)$ and $\kappa = \lambda _{D}^{-1}$ is the inverse Debye screening length. This can provide a discernible flow enhancement in relatively concentrated solutions of small $\lambda _{D}$~\cite{bouzigues.ci:2008}, but not in dilute solutions where $\lambda _{D}$ is of the order of hundreds of nm. Eq.\eqref{eq:hydrophobic} implies that adsorbed at the slippery surface charges are immobile, but if they move in response to the field, its second term could be significantly reduced~\cite{maduar.sr:2015,silkina.ef:2019}. Eq.\eqref{eq:hydrophobic} with $\kappa b \gg 1$ suggests that a massive amplification of electro-osmotic flow can be achieved over super-hydrophobic (Cassie) surfaces with trapped gas bubbles that significantly enhance the hydrodynamic slip~\cite{joseph.p:2006,ou.j:2005,nizkaya.tv:2016}. During the last decade several authors concluded that this is, indeed, possible, but only with charged
liquid-gas interfaces~\cite{bahga.ss:2009,squires.tm:2008,belyaev.av:2011a}.

Another avenue could be to employ charged porous coatings that are permeable to water and ions, such as polyelectrolyte networks, ion-exchange resins, silica gels, porous membranes, polyelectrolyte multilayers and brushes, and, in fact, any hydrophilic microtextured  surfaces in the Wenzel state. \citet{brinkman.hc:1949} proposed a modification of the Darcy law to accommodate situations involving shear rates of an outer fluid at the surface of the porous medium. This led to the generalized Stokes equation for a (tangent) pressure-driven flow and to the concept of the Brinkman screening length, $\Lambda$, defined as the square root of  the hydrodynamic (Darcy) permeability of the medium. Such an approach is widely employed in hydrodynamics of porous media leading, in particular, to the prediction of the exponential decay of a fluid velocity at the permeable surface to the (finite) Darcy velocity inside the porous medium.
As proposed by \citet{beavers.gs:1967}, this velocity drop is proportional to the shear rate of an outer fluid and can be very large~\cite{gupte.sk:1997}.
Despite the obvious importance of an emerging liquid slip at the porous surface for a potential amplification of electroosmotic flows, fundamental understanding of this did not begin to emerge until quite recently. In general, whilst considerable progress has been made over the last decades in understanding the equilibrium properties of porous surfaces in electrolyte solutions, their electro-hydrodynamic   properties are relatively less well understood. There is some literature describing attempts to provide a satisfactory theory of electro-osmosis near porous surfaces. We mention below what we believe are the more relevant contributions.

\citet{donath.e:1979} appear to have been the first to study theoretically the  electro-osmotic velocity near a porous permeable film. These authors addressed themselves the case of low potentials and calculated the velocity by including into the Stokes equation inside the film the so-called `friction coefficient', which is  obviously equivalent to  $\Lambda^2$. One of the main results of this pioneering work is that $\Psi_s$ of porous surfaces does not define unambiguously the flow properties. As a consequence,  $V_{\infty}$ does not vanish at high salt concentration, where $\Psi_s \simeq 0$, as it would be for impermeable walls. The authors, however, failed to propose a physical interpretation of these results.

A more systematic treatment of the influence of the `friction coefficient' on electro-osmosis was contained in a paper published by  \citet{ohshima.h:1990a}.  These authors relaxed the low potential assumption and proposed an expression relating $V_{\infty}$ to the integral of an electro-static potential. However, they concluded that a substitution of the potential profile to this expression give `results too complex for practical use', so that the case of only $\Lambda/H \ll 1$, where $H$ is the thickness of the porous film, was resolved explicitly. For this situation \citet{ohshima.h:1990a} predicted that $V_{\infty}$ is controlled, besides $\Psi_s$, by the Brinkman and inner Debye screening lengths, and also by the Donnan potential of the porous medium, but did not present any  calculations illustrating or verifying their theoretical results. Similar remark applies to a paper by \citet{ohshima.h:1995} that, although about electrophoretic properties of spherical particles with porous coatings, is directly relevant to electro-osmosis when concerns a situation of their very large radius. The author has generalized his prior work~\cite{ohshima.h:1990a} to several configurations, such as, for example, a charged wall and a neutral coating,  and generally confirmed an earlier conclusion about finite $V_{\infty}$ at high salt concentrations~\cite{donath.e:1979}, but did not attempt to relate his results to the inner flow and emerging liquid velocity at the interface.

Subsequent attempts at improvement the description of electro-osmotic flow near porous surfaces have been concerned mostly the lifting of the standard assumption of the uniform volume charge density and `friction coefficient' for some specific polymer systems. \citet{duval.jfl:2004,duval.jfl:2005} introduced a concept of the `diffuse soft layer', where both parameters  decrease linearly from a constant `bulk' values to zero, and presented a solution for velocities, which is expressed by infinite series. These authors argue that such a model improves the fit of the low salt data, compared to prior theoretical work~\cite{ohshima.h:1990a,ohshima.h:1995}, although they
have ignored,  that experimentalists often used a linear,  overestimating the electrostatic potential, version of the theory, which is unsuitable for dilute solutions. Consequently, the questions whether and how  should one take into account the possible non-uniformity of the porous material near an interface, remain open, but we consider papers~\cite{duval.jfl:2004,duval.jfl:2005} to be important contributions.

While existing theories of electro-osmosis near porous systems~\cite{donath.e:1979,ohshima.h:1990a,ohshima.h:1995,duval.jfl:2004,duval.jfl:2005,ohshima.h:2006} are frequently invoked in the interpretation of the electro-osmotic velocity, its relation to zeta potential has remained somewhat obscure. Following \citet{smoluchowski.m:1921} some authors~\cite{ohshima.h:1990a,ohshima.h:1995,duval.jfl:2004} termed the static potential at the location of the no-slip boundary condition `a zeta-potential' and concluded that it `loses its significance'~\cite{ohshima.h:1990a} or `is undefined and thus nonapplicable'~\cite{duval.jfl:2004}. However, such a definition of $Z$ is a consequence of the no-slip postulate that is unsuitable for porous surfaces. Although the authors do not recognize this, their results simply imply that for permeable interfaces $Z \geq \Psi_s$ (which is equivalent to $\mathcal{A} \geq 1$). More recent calculations also indicate that $Z$ typically exceeds $\Psi_s$~\cite{sobolev.vd:2017,chen.g:2015}.
 Neither paper addresses itself to the issues of surface slip. This was taken up only recently in the paper by \citet{silkina.ef:2020b}, who carried out calculations in the limit of infinite Brinkman length, $\Lambda \to \infty$,
in an attempt to obtain a proper understanding of an upper bound on achievable zeta-potential. These authors concluded that  $Z$ of a porous surface can potentially exhibit an enhancement by an order of magnitude or more due to the emergence of a large surface slip,  thanks to the fluid flow in a porous film enriched by counter-ions. Nevertheless, general principles to control $Z$ of porous surfaces of a finite $\Lambda$ are not established yet, and we also gained the impression that several crucial aspects of the electro-hydrodynamics of the porous interface have been given so far insufficient attention.

In the present paper, we give a general theoretical description of the electroosmosis near porous surfaces with the focus on the zeta-potential and flow amplification. We consider
a planar, uniformly charged porous films placed in contact with a reservoir of electrolyte solution. Our analytic theory provides an explanation of the
variation of electro-osmotic velocity with the volume charge density of the coating, ionic
concentration, and Brinkman length. It being very well suited  to exact calculations replacing numerical work, as well as  has the merit of yielding useful (approximate) analytical results in some limits. These features are especially advantageous when  one is attempting to calculate velocities at  arbitrary values of parameters. Our results provide new insight into the physics of electro-osmosis, zeta-potential, and the nature of electro-osmotic slip at the porous interface.

Our paper is arranged as follows. In Sec.~\ref{sec:general} some general considerations concerning velocities of a liquid, zeta-potential, and flow amplification are presented. The summary of theoretical relationships for an electrostatic potential is given, and the concept of an electro-osmotic slip length is introduced. Sec.~\ref{Surface_slip} describes theoretical results. In Sec.~\ref{kineqK} the case of $\kappa _{i} \neq
\mathcal{K}$, where $\kappa _{i}$ is the inner screening length and $\mathcal{K} = \Lambda^{-1}$, is described and the exact and approximate equations for the slip velocity, zeta-potential, amplification factor, and slip length are derived. The analogous results for the case of $\kappa _{i} =
\mathcal{K}$ are presented in Sec.~\ref{kieqK}. Sec.~\ref{sec:results} contains the results of numerical calculations of the velocity profiles in both large and small Brinkman length situations, validating the theoretical predictions. Numerical and approximate theoretical results for zeta-potentials are presented and compared with surface potentials and velocities at the surface. Finally, the amplification factor and slip length are discussed and contrasted. The issues of implications of these results and the applicability range of our asymptotic approximations are also addressed. In Sec.~\ref{sec:salt} results for the zeta-potential and slip length, plotted vs. salt concentration, are presented for specific porous films.
We
 conclude in Sec.~\ref{sec:conclusion} with a further discussion of our results and their possible relevance for electro-kinetic experiments.

\section{General considerations}\label{sec:general}

\begin{figure}[tbp]
\begin{center}
\includegraphics[width=9cm]{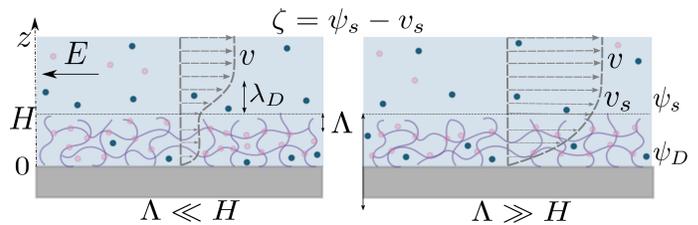}
\end{center}
\par
\vspace{-0.4cm}
\caption{Charged porous film of thickness $H$ in contact with an electrolyte
solution. Anions and cations are denoted with
bright and dark circles. The film is permeable for ions and solvent, so that the Donnan, $%
\protect\psi_D$, and surface, $\protect\psi_s$, static potentials are
established self-consistently. An outer electrostatic diffuse
layer of a thickness, which is of the order of Debye screening length, $%
\protect\lambda_D$, is formed in the vicinity of
the porous film. The application of a tangential electric field, $E$, leads
to an electro-osmotic flow of a solvent (shown by arrows) that depends on the Brinkman screening length $\Lambda$. Due to emerging velocity at the surface its dynamic zeta-potential $\zeta \geq \psi_s.$}
\label{Fig:sketch}
\end{figure}

The system geometry is shown in Fig.~\ref{Fig:sketch}. Rather  than  trying  to  solve the problem at the scale of the individual pores, it is appropriate to consider the `macroscale' situation  of the imaginary smooth and homogeneous coating that mimics the actual (porous heterogeneous) one and has the same effective properties. We, thus, consider the homogeneous permeable film of a thickness $H$ (that will be the reference length scale of our problem) and a fixed volume charge density $\varrho$, taken positive without loss of generality, on a solid support. The film is in contact with a semi-infinite 1:1 electrolyte of
concentration $c_{\infty}$, permittivity $\varepsilon$, and dynamic
viscosity $\eta$.
Ions obey Boltzmann distribution, $c_{\pm
}(z)=c_{\infty}\exp (\mp \psi (z))$, where $\psi (z)=\textsl{e} \Psi (z)/(k_{B}T)$ is
the dimensionless electrostatic potential, $\textsl{e}$ is the elementary positive
charge, $k_{B}$ is the Boltzmann constant, $T$ is a temperature, and the
upper (lower) sign corresponds to the cations (anions). The inverse Debye
screening length of an electrolyte solution, $\kappa \equiv \lambda_D^{-1}$,
is defined as usually, $\kappa^2=8 \pi \ell_B c_{\infty}$, with the Bjerrum
length $\ell_B=\dfrac{\textsl{e}^2}{\varepsilon k_BT}$. The Debye length defines a new (electrostatic) length scale and is the measure of the thickness of the outer diffuse layer.

Ions of an outer electrolyte can permeate inside the porous film, giving rise to their homogeneous equilibrium distribution in the system, with the enrichment of anions in the film. We consider here only thick, compared to the inner diffuse layer, films, with an extended `bulk' electro-neutral region (where intrinsic coating charge is completely screened by absorbed electrolyte ions, is formed). The potential in this region is referred below to as the Donnan potential, $\psi_D$. The surface potential $\psi_s$
is defined at $z = H$.

The system subjects to a weak tangential electric field $E$, so that in steady state $\psi (z)$ is independent of the fluid flow and satisfies the nonlinear Poisson-Boltzmann equation
\begin{equation}
\psi_{i, o}^{^{\prime\prime}} = \kappa^2 \left(\sinh\psi_{i, o} -\rho\Theta
\left(H - z \right)\right) ,  \label{Eq:PB_io}
\end{equation}
where $^{\prime}$ denotes $d/d z$, with the index $\{i, o\}$ standing for
``in'' $(z \leq H)$ and ``out'' $(z \geq H)$,  $\Theta(z)$ is the Heaviside step function, and $\rho = \dfrac{\varrho}{%
2 \textsl{e} c_{\infty}}$. We will term films of $\rho \ll 1$ weakly charged, and those of $\rho \gg 1$ highly charged.

Note that Eq.\eqref{Eq:PB_io} immediately suggests that
\begin{equation}
\psi_D \equiv \psi _{0} =\arsinh(\rho ),  \label{eq:phi_0_thick}
\end{equation}%
where $\psi_0 = \psi_i (0)$, since in the electro-neutral area $\psi_{i}^{^{\prime\prime}}$ vanishes.
The potential drops in the inner diffuse layer as
\begin{equation}
\psi _{i}(z)\simeq \psi _{D} - \Delta \psi e^{\kappa _{i}(z-H)}
\label{eq:phi_in_thick2}
\end{equation}%
Eq.\eqref{eq:phi_in_thick2} was first derived for weakly charged films by~\citet{ohshima.h:1985}, and recently strictly justified for coatings of any $\rho$~\cite{silkina.ef:2020b}. Here $\Delta \psi= \psi_D - \psi_s$ depends on $\rho$ only~\cite{silkina.ef:2020}
\begin{equation}  \label{eq:psi_s_thick}
\Delta \psi  = \frac{\cosh \psi_D - 1}{\rho} = \dfrac{\sqrt{1+\rho^2} - 1}{\rho}
\end{equation}
and the inner screening length
\begin{equation}\label{eq:kappa_i}
\kappa _{i}=\kappa (\cosh {\psi _{D}})^{1/2} = \kappa
(1+\rho ^{2})^{1/4}
\end{equation}%
is the function of both $\kappa$ and $\rho$. The criterion of a thick film we consider here is then
\begin{equation}\label{eq:criterion}
   \kappa (1+\rho ^{2})^{1/4} H  \gg 1
\end{equation}

In the outer diffuse layer the potential decays from
\begin{equation}\label{eq:psi_s_thick1}
\psi_s = \arsinh(\rho) + \displaystyle\frac{1-\sqrt{1+\rho^2}}{\rho}
\end{equation}
 down to zero in the bulk electrolyte as~\cite{andelman.d:2006book}
\begin{equation}  \label{eq:PB_out3}
\psi_{o} (z) = 4 \artanh \left[ \gamma e^{-\kappa (z-H)}\right],
\end{equation}
with $\gamma =\tanh\dfrac{\psi_s}{4}$.

In the limits of small and large $\rho$,  the above equations can be simplified. When $\rho \ll 1$, Eqs.\eqref{eq:psi_s_thick}, \eqref{eq:kappa_i}, and \eqref{eq:psi_s_thick1} reduce to~\cite{ohshima.h:1985}
\begin{equation}\label{eq:small_rho_summary}
 \Delta \psi \simeq \dfrac{\rho}{2}, \, \kappa_i \simeq \kappa, \, \psi_s  \simeq  \dfrac{\rho}{2}
\end{equation}
If $\rho \gg 1$ they transform to~\cite{silkina.ef:2020}
\begin{equation}\label{eq:large_rho_summary}
  \Delta \psi \simeq 1, \, \kappa_i \simeq \kappa \sqrt{\rho}, \, \psi_s \simeq  \ln (2\rho) - 1,
\end{equation}

For our geometry the concentration gradients at every location are perpendicular to  the direction of the flow, it is therefore legitimate to  neglect advection. Consequently, the dimensionless velocity of an electro-osmotic flow, $%
v(z)=\displaystyle\frac{4\pi \ell _{B}\eta }{\textsl{e} E}V(z)$, satisfies the generalized Stokes equation
\begin{equation}
v_{i,o}^{\prime \prime }-\mathcal{K}^{2}  v_{i,o} \Theta (H-z)=\psi _{i,o}^{\prime
\prime }+\kappa ^{2}\rho \Theta (H-z),  \label{eq:Stokes}
\end{equation}%
where $\mathcal{K} = \Lambda^{-1}$ is
the inverse Brinkman length. The equation contains
multiple length scales due to the different physical
effects involved.

At the wall we apply a classical no-slip
condition, $v_{0}=v_i(0)=0$, and at the surface the condition of continuity of velocity, $v_i (H) = v_o (H)$, and shear rate, $v_i^{\prime } (H) = v_o^{\prime } (H)$, is imposed.
Far from the surface, $z \to \infty$, the solution of Eq.\eqref{eq:Stokes} should satisfy $v_o^{\prime} (\infty) = 0$ to provide a plug flow.
Its constant, i.e. independent on $z$, velocity is denoted below as $ v_{\infty}$.

The outer $v$-profile is given by
\begin{equation}  \label{eq:v_o}
v_{o}(z) = v_{\infty} + \psi_{o}(z),
\end{equation}
where $\psi_{o}$ is defined by Eq.\eqref{eq:PB_out3}, and
\begin{equation}  \label{eq:v_inf}
 v_{\infty} = v_s - \psi_{s} = -
\zeta,
\end{equation}
where
$v_s = v(H)$ is the
liquid velocity at surface and $\zeta = \textsl{e} Z / (k_B T) $ is the dimensionless zeta-potential. Note that it follows from Eq.(\ref{eq:v_inf}) that the enhanced electro-osmotic mobility is expected due to large $\psi_s$, which is an equilibrium property of the system, as well as large $v_s$ that reflects a hydrodynamic permeability of the porous coating. The amplification factor can be then expressed as
\begin{equation}\label{eq:amplification}
 \mathcal{A} =   \frac{\zeta}{\psi_s} = 1-  \frac{v_s}{\psi_s}
\end{equation}
The outer problem, thus,
reduces to calculation of (negative) $v_s$.

The velocity jump inside the porous film, $\Delta v = v_0 - v_s = - v_s$. In common applications, the film is much thinner than any of the
macroscopic dimensions. Therefore, the liquid appears to slip at the velocity $-v_s$ along
the surface of a porous coating. It is then possible to define an effective (positive definite) electro-osmotic slip length using the boundary condition $-\Delta v = b v_o^{\prime }(H)$, where the outer liquid  velocity is proportional to the shear strain rate via the slip length. In such a definition $b$ is the distance from the surface at which the outer flow profile extrapolates to zero.
From Eq.\eqref{eq:v_o} it follows that $v_o^{\prime }(H) = \psi_{o}' (H)$. Therefore, the shear rate is
\begin{equation}\label{eq:PB_out2}
v_{o}' (H) =-2\kappa \sinh \left(\frac{\psi_s}{2}\right),
\end{equation}
and we obtain
\begin{equation}\label{eq:slip_length}
b \kappa = - \dfrac{v_s}{2 \sinh \left(\dfrac{\psi_s}{2}\right)}
\end{equation}
Substituting $v_s$ from the latter relation into Eq.\eqref{eq:amplification} we recover Eq.\eqref{eq:hydrophobic}. This indicates that an electro-osmotic flow near porous films is identical to that near slippery impermeable surfaces with immobile surface charges, but $\kappa b$ and $\psi_s$ are established self-consistently. Using $\psi_s$ given by Eqs.\eqref{eq:small_rho_summary} and \eqref{eq:large_rho_summary} it is possible to obtain sensible approximations in the limits of weakly and highly charged films. When $\rho \ll 1$
\begin{equation}\label{eq:slip_length_small_rho}
b \kappa \simeq - \dfrac{v_s}{\psi_s} \simeq - \dfrac{2 v_s}{\rho},
\end{equation}
and for $\rho \gg 1$ standard manipulations yield
\begin{equation}\label{eq:slip_length_large_rho}
b \kappa \simeq  - \sqrt{\dfrac{e}{2 \rho}} v_s,
\end{equation}
where $e$ is the base of the natural logarithm.

\section{Surface slip, zeta-potential, and flow amplification}\label{Surface_slip}

In order to obtain a detailed information concerning zeta-potential, outer flow amplification and slip length a calculation of $v_s$ arising due to the inner flow is required. We have obtained the inner velocity profiles by solving Eq.(\ref{eq:Stokes}) with $\psi_i$ satisfying Eq.(\ref{eq:phi_in_thick2}) and prescribed boundary conditions. Below we consider two distinct cases, of $\kappa _{i} \neq
\mathcal{K}$ and $\kappa _{i} =
\mathcal{K}$, that lead to different forms of the solution for  $v_i$ and $v_s$.

\subsection{The case of $\kappa _{i} \neq
\mathcal{K}$}\label{kineqK}

In these circumstances

\begin{widetext}
	%\centering
\begin{equation}
v_{i}=C_{1}\left( e^{-\mathcal{K}z}-1\right) +C_{2}\left( e^{\kappa
_{i}z}-e^{-\mathcal{K}z}\right) +C_{3}\sinh \left( \mathcal{K}z\right) ,
\label{eq:EO_in_thick}
\end{equation}

%\end{widetext}
where

\begin{equation}\label{eq:C1C2}
C_{1}=\rho \left( \frac{\kappa }{\mathcal{K}}\right) ^{2},\
C_{2}=-\frac{\kappa _{i}^{2}e^{-\kappa _{i}H}\Delta \psi }{\kappa _{i}^{2}-\mathcal{K}^{2}},
 \end{equation}%
  \begin{equation}\label{eq:C3}
 C_{3} =\frac{1}{\cosh\mathcal{K}H} \left[ \frac{\kappa_i \mathcal{K} \Delta \psi}{\kappa_i^{2}-\mathcal{K}^{2}}+(C_1-C_2)e^{-\mathcal{K}H} \right]
\end{equation}%
Note that Eq.\eqref{eq:EO_in_thick} is invalid when $\kappa_i =\mathcal{K}$ since both $C_2$ and $C_3$ diverge. This special case should be treated separately, and we will return to this point in Sec.\ref{kieqK}.
Eqs.\eqref{eq:EO_in_thick}-\eqref{eq:C3} allow us to obtain the following expression for an emerging slip velocity

%\begin{widetext}
	%\centering
\begin{equation} \label{eq:v_s_thick}
v_s=\frac{\kappa_i^2 \Delta \psi}{\kappa_i^{2}-\mathcal{K}^{2}}\left(\dfrac{\mathcal{K}}{\kappa_i} \tanh \mathcal{K}H - 1 \right)+\left[\rho \left( \frac{\kappa }{\mathcal{K}}\right) ^{2}+\frac{\kappa _{i}^{2}e^{-\kappa _{i}H}\Delta \psi }{\kappa _{i}^{2}-\mathcal{K}^{2}}\right](1+\tanh \mathcal{K}H)e^{-\mathcal{K}H}-\rho \left( \frac{\kappa }{\mathcal{K}}\right) ^{2},
\end{equation}%
\end{widetext}
where the potential drop in the film $\Delta \psi$ is given by Eq.\eqref{eq:psi_s_thick} and $\kappa _{i}$ is expressed by Eq.\eqref{eq:kappa_i}. Eq.\eqref{eq:v_s_thick} can be used for thick films of any $\rho$ and $\mathcal{K}H$. As follows from Eq.\eqref{eq:v_inf}, once $v_s$ is determined, the zeta potential can be immediately obtained using $\zeta =  \psi_s - v_s$, with the surface potential expressed by Eq.\eqref{eq:psi_s_thick1}.

Eq.\eqref{eq:v_s_thick} can be simplified in the limits of large and small Brinkman length. Below we discuss these two situations.

\subsubsection{Large Brinkman length ($\mathcal{K}H\ll 1$)}\label{LBL}

We first consider a somewhat idealized case of the very large Brinkman length $\Lambda$, where an additional dissipation in the porous film can be neglected. This can be seen as an upper  bound on the electro-osmotic velocity that constrains its attainable (largest)
value.

For small $\mathcal{K}H$ the inner velocity given by Eq.(\ref{eq:EO_in_thick}) can be expanded in series for small $\mathcal{K}H$ and $\mathcal{K}z$, and to leading  order
\begin{equation} \label{eq:EO_in_large_thick}
v_{i}=-\Delta \psi e^{\kappa _{i}(z-H)}-\rho \kappa ^{2}\left( Hz-\frac{z^{2}%
}{2}\right) +O\left( \mathcal{K}z\right),
\end{equation}%
The first term is associated with the reduction of the potential in the inner diffuse layer, and is equal to $-(\psi_D - \psi_i)$ (see Eq.\eqref{eq:phi_in_thick2}). This  contribution to $-v_{i}$ decreases exponentially from 0 to $\Delta \psi$ with increasing $z/H$ from 0 to 1.  The second term is associated with a body force $\rho \kappa^2$ that drives the inner flow (see Eq.\eqref{eq:Stokes}) by acting on the mobile ions accumulated in the Donnan portion of the film. This contribution resembles  the usual no-slip parabolic Poiseuille flow. One can define a hydrodynamic permeability of such a film as a ratio of the flow rate (expressed per unit film width) and the driving force. Performing integration from 0 to $H$ of the expression in brackets in Eq.\eqref{eq:EO_in_large_thick} and dividing by $H$ we find that the hydrodynamic permeability of the film  is equal to  $H^2/3$, i.e.  is varying as the square of its thickness, but does not depend on the Brinkman length.

It follows from Eq.\eqref{eq:EO_in_large_thick} that
\begin{equation}  \label{eq:v_s_large}
v_s \simeq - \Delta \psi - \frac{\rho (\kappa H)^{2}}{2}
\end{equation}
Since $\Delta \psi \leq 1$, the second term should dominate  even at moderate $\rho$.

For weakly charged films, $\rho\ll 1$, the last equation is reduced to
\begin{equation}  \label{eq:v_s_large_smallrho}
v_s \simeq - \frac{\rho}{2} (1 + (\kappa H)^{2}),
\end{equation}
obtained using \eqref{eq:small_rho_summary}. From Eqs.\eqref{eq:small_rho_summary}, \eqref{eq:v_inf} and \eqref{eq:amplification} it follows that

\begin{equation}\label{eq:zeta1}
  \zeta \simeq \frac{\rho}{2} (2 + (\kappa H)^{2}),
\end{equation}
and

\begin{equation}\label{eq:zeta_large_rholl1}
\mathcal{A} \simeq 2 + (\kappa H)^2
\end{equation}
Substituting $v_s$ given by Eq.\eqref{eq:v_s_large_smallrho} into \eqref{eq:slip_length_small_rho} we obtain
\begin{equation}\label{eq:b_large_rholl1}
\kappa b \simeq 1 + (\kappa H)^2
\end{equation}
We remark that in this low $\rho$ regime $\mathcal{A}$ and $\kappa b $ do not depend on $\rho$, and are finite even if $\rho \to 0$, where $\psi_s \simeq 0$.  At first sight this is somewhat surprising, but we recall that our dimensionless charge density is introduced by dividing the real one by the salt concentration, so that a nearly vanishing $\rho$ simply implies that the film is enriched by counter-ions that practically fully screen its intrinsic charge. It then becomes almost self-evident that these absorbed mobile ions should induce some inner flow and slip velocity.
Indeed, Eqs.\eqref{eq:v_s_large_smallrho} and \eqref{eq:zeta1} predict that large slip velocity and zeta-potential can be generated even by relatively weakly charged thick films, i.e. when the impact of diffuse layers on the electro-osmosis is marginal. This suggests that an outer flow takes it origin  mostly in the `bulk' portion of the coating, where the total electroneutrality condition is satisfied.
It is tempting to speculate that one can significantly amplify $v_{\infty}$ making porous film thicker. However, when the film becomes thick enough, the condition $\mathcal{K} H \ll 1$ violates, and Eq.\eqref{eq:EO_in_large_thick} is no longer valid.

For highly charged films, $\rho\gg1$, using \eqref{eq:large_rho_summary} we find
\begin{equation}  \label{eq:v_s_large_largerho}
v_s \simeq -1 -  \frac{\rho (\kappa H)^2}{2},
\end{equation}
The corresponding zeta-potential and flow amplification can then be found from Eqs.\eqref{eq:large_rho_summary}, \eqref{eq:v_inf} and \eqref{eq:amplification}
 \begin{equation}\label{eq:zeta2}
  \zeta \simeq \ln(2\rho) + \frac{\rho (\kappa H)^2}{2},
\end{equation}
\begin{equation}\label{eq:zeta_large_rhogg1}
\mathcal{A} \simeq  1 + \dfrac{2 + \rho (\kappa H)^{2}}{2 (\ln(2\rho)-1)}
\end{equation}
Using Eqs.\eqref{eq:slip_length_large_rho} and \eqref{eq:v_s_large_largerho} we argue that $\kappa b$ can be approximated by
\begin{equation}\label{eq:b_large_rhogg1}
  \kappa b \simeq \dfrac{ (\kappa H)^2 \sqrt{2 e \rho}}{4}
\end{equation}

Note that in this limit of high intrinsic volume charge density the functions $\zeta$, $\mathcal{A}$, and $\kappa b$ depend on $\rho$ and $\kappa H$, and these dependences are nonlinear. Another important remark would be that the contributions of $\rho$ and $\kappa H$ are decoupled.

\subsubsection{Small Brinkman length ($\mathcal{K}H\gg 1$)}\label{SBL}

 Let us now consider the limit of small Brinkman length or $\mathcal{K} H \gg 1$. It is generally accepted that in such a situation the slip velocity nearly vanishes, and $\psi_{s} \simeq \zeta$ with $\mathcal{A} \simeq  1$. This limit is, therefore, often seen as a lower bound on
the electro-osmotic velocity that constrains its attainable minimal value~\cite{silkina.ef:2020b}. Below we demonstrate that, despite a common belief, in some situations one can generate a strong flow even with small $\Lambda$.

We first note that in this situation
$C_3$ given by Eq.\eqref{eq:C3} reduces to
\begin{equation} \label{eq:C3_thick_betagg1}
C_{3}= 2 e^{-\mathcal{K}H}\frac{\kappa _{i}\mathcal{K} }{\kappa _{i}^{2}-%
\mathcal{K}^{2}} \Delta \psi,
\end{equation}%
The $v_i$-profile
\begin{widetext}
	\centering
\begin{equation}\label{eq:SBL_vi}
v_{i}= -\frac{\kappa _{i}\Delta \psi }{\kappa _{i}^{2}-\mathcal{K}^{2}%
}\left( \kappa _{i}e^{\kappa _{i}\left( z-H\right) }-\mathcal{K}e^{\mathcal{K%
}\left( z-H\right) }\right) - \rho \left( \frac{\kappa }{\mathcal{K}}\right) ^{2}\left(1 - e^{-\mathcal{
K}z}\right)
\end{equation}%
\end{widetext}
can then be easily ascertained to find
\begin{equation} \label{eq:v_s_thick_betagg1}
v_{s} \simeq -\frac{\kappa _{i} }{\kappa _{i}+\mathcal{K}} \Delta \psi -\rho
\left( \frac{\kappa }{\mathcal{K}}\right) ^{2}.
\end{equation}
Eq.\eqref{eq:v_s_thick_betagg1} indicates that $v_s$ is a superposition of a flow in the inner diffuse layer that is linear in $\Delta \psi$, but now also depends on
$\kappa _{i}/\mathcal{K}$, and of a plug flow emerging in the Donnan region. The later, naturally, satisfies Eq.\eqref{eq:Stokes} when $\psi_{i}^{^{\prime\prime}} = 0$ and $v_i^{^{\prime\prime}} = 0$
\begin{equation}\label{eq:plateau0}
  v_i =  -\rho \left( \frac{\kappa }{\mathcal{K}}\right) ^{2} \equiv -\rho  \varkappa^2,
\end{equation}
where we have introduced
\begin{equation}\label{eq:varkappa}
  \varkappa = \dfrac{\kappa }{\mathcal{K}} \equiv \dfrac{\Lambda}{\lambda_D}
\end{equation}
From Eq.\eqref{eq:plateau0} one can easily calculate the hydrodynamic permeability defined in Sec.~\ref{LBL}, and obtain that it is equal to  $\Lambda^2$, i.e. coincides with the Darcy permeability of the porous medium in a pressure-driven flow.

Two limits can now be distinguished depending on the value of $\rho$ as we did in Sec.\ref{LBL}.

The limit of small
volume charge density, $\rho \ll 1$, yields

\begin{equation}\label{eq:v_s_large_beta_rholl1}
v_{s} \simeq -\frac{\rho \varkappa}{2(\varkappa +1)}  -\rho \varkappa^{2},
\end{equation}
obtained using Eqs.\eqref{eq:small_rho_summary}.
Using Eqs.\eqref{eq:small_rho_summary}, \eqref{eq:v_inf} and \eqref{eq:amplification} we then find

\begin{equation}\label{eq:zeta3}
  \zeta \simeq \frac{\rho}{2} \left(1 + \frac{\varkappa}{\varkappa +1}  + 2 \varkappa^{2}\right),
\end{equation}

\begin{equation}\label{eq:A_small_rholl1}
\mathcal{A} \simeq 1 + \frac{\varkappa}{\varkappa +1}  + 2 \varkappa^{2},
\end{equation}
and Eq.\eqref{eq:slip_length_small_rho} gives
\begin{equation}\label{eq:b_small_rholl1}
\kappa b \simeq 2 \varkappa^{2} + \frac{\varkappa}{1 + \varkappa}
\end{equation}

When $\varkappa \ll 1$, $v_{s} \simeq -\rho \varkappa /2,$ i.e. nearly vanishes, $\kappa b \simeq \varkappa$, $\mathcal{A} \simeq 1$, and $\psi
_{s}\simeq \zeta \simeq \rho/2 \ll 1$. Thus the electro-osmotic flow is equivalent to expected for impermeable surfaces of the same $\psi_s$.
However, when $\varkappa \gg 1$, $v_{s} \simeq -\rho \varkappa^{2}$, $\mathcal{A} \simeq \kappa b \simeq 2 \varkappa^{2}$, and $\zeta \simeq \rho \varkappa^{2} \simeq - v_s$. The overall conclusion from these is that $\zeta$ can be quite large
despite small $\rho$, and reflects mostly $v_s$, but not the surface potential. The main contribution to $\zeta$ comes from
the Donnan region, where the potential (as well as
the ionic concentrations are almost constant),  but not from the inner diffuse layer.

Repeating the above calculations for the limit of $\rho \gg 1$ (using Eqs.\eqref{eq:large_rho_summary} and \eqref{eq:slip_length_large_rho} instead of Eqs.\eqref{eq:small_rho_summary} and \eqref{eq:slip_length_small_rho}) yields different approximate expressions

\begin{equation} \label{eq:v_s_large_beta_rhogg1}
v_{s} \simeq -\frac{\varkappa \sqrt{\rho}}{1 + \varkappa \sqrt{\rho}} -\rho \varkappa^{2},
\end{equation}

\begin{equation} \label{eq:v_bulk_large_beta_rhogg1}
\zeta \simeq \frac{\varkappa \sqrt{\rho}}{1 + \varkappa \sqrt{\rho}} + \rho \varkappa^{2} + \ln (2 \rho) - 1,
\end{equation}

\begin{equation}\label{eq:A_large_beta_rhogg1}
\mathcal{A} \simeq 1 + \dfrac{\dfrac{\varkappa \sqrt{\rho}}{1 + \varkappa \sqrt{\rho}} + \rho \varkappa^{2}}{\ln (2 \rho) - 1},
\end{equation}

\begin{equation}\label{eq:b_small_rhogg1}
\kappa b \simeq \varkappa \sqrt{\frac{e}{2}} \left[ \frac{1}{1 + \varkappa \sqrt{\rho}} + \varkappa \sqrt{\rho}  \right]
\end{equation}
We see that in this regime the electroosmosis is controlled by $\varkappa \sqrt{\rho}$. For small $\varkappa \sqrt{\rho}$ we get $v_s \simeq -\varkappa \sqrt{\rho} \simeq 0$,  $\zeta \simeq \psi_s$, $\mathcal{A} \simeq 1$, and $\kappa b \simeq \varkappa \sqrt{\dfrac{e}{2}}$. The sole role of a porous film is thus to set $\psi_s$.

When $\varkappa \sqrt{\rho} \gg 1$, the large surface slip $v_s \simeq - \rho \varkappa^2$ and zeta-potential $\zeta \simeq \rho \varkappa^{2} + \ln (2 \rho)$ are generated. We remark, that both $v_s$ and $\psi_s$ contribute to the value of $\zeta$. The amplification factor $\mathcal{A} \simeq 1 + \dfrac{ \rho \varkappa^{2}}{\ln (2 \rho)-1}$, indicating that the flow is significantly enhanced. Finally, $\kappa b \simeq \varkappa^2 \sqrt{\dfrac{e \rho}{2}}$. Overall we conclude that one can generate a very strong electro-osmotic flow even when the Brinkman length is small, provided both $\rho$ and $\varkappa \sqrt{\rho}$  are large, i.e. when the  Brinkman length of the highly charged film is larger than the Debye screening length.

\subsection{The case of $\kappa _{i} =
\mathcal{K}$}\label{kieqK}

When $\kappa_i=\mathcal{K}$
the inner velocity is given by
\begin{widetext}
	\centering
	\begin{equation} \label{eq:EO_in_thick_kiH=beta}
v_{i}= -\frac{\Delta\psi}{2} \left( \kappa_i z e^{\kappa_i (z-H)} -\frac{(\kappa_i H-1 )}{\cosh (\kappa_i H)}\sinh \left(\kappa_i z\right)\right) - \rho \left( \frac{\kappa }{\mathcal{K}}\right) ^{2}\left(1- e^{- \kappa_i z}\right),
\end{equation}
\end{widetext}
and slip velocity at a thick film is then equal to
\begin{equation} \label{eq:EO_v_s_thick_kiH=beta}
v_{s}=-\frac{\Delta \psi }{2}-\rho \left( \frac{\kappa }{\mathcal{K}}\right) ^{2}
\end{equation}%
Using Eqs.\eqref{eq:kappa_i} and \eqref{eq:varkappa} the second term can be reformulated as
\begin{equation}\label{eq:EO_v_s_thick_kiH=beta1}
\rho \varkappa^{2} =  \dfrac{\rho}{\sqrt{1 + \rho^2}} \leq 1,
\end{equation}
i.e. slip velocity depends on $\rho$ only and $-v_s \leq 3/2$.

Employing the same arguments as in Sec.~\ref{kineqK} we find approximations for weakly and highly charged films. In the limit of
$\rho \ll 1$ both the slip velocity and zeta-potential are linear in $\rho$
\begin{equation}\label{eq:v_ki_rholl1}
v_{s} \simeq -\frac{5 \rho}{4}, \,
  \zeta \simeq \frac{7 \rho}{4},
\end{equation}
but the amplification factor and $\kappa b$ are constant
\begin{equation}\label{eq:A_ki_rholl1}
\mathcal{A} \simeq \frac{7}{2}, \,
\kappa b \simeq \frac{5}{2}
\end{equation}
If $\rho \gg 1$, a maximum possible slip velocity for the case $\kappa_i=\mathcal{K}$ is reached and we derive that $\zeta$, showing a weak logarithmic growth with $\rho$, reflects mostly $\psi_s$
\begin{equation}\label{eq:v_ki_rhogg1}
v_{s} \simeq - \dfrac{3}{2}, \, \zeta \simeq  \dfrac{1}{2}  + \ln (2 \rho)
\end{equation}
Amplification factor and $\kappa b$ decrease with $\rho$
\begin{equation}\label{eq:A_ki_rhogg1}
\mathcal{A} \simeq 1 + \dfrac{3}{2 ( \ln (2 \rho) - 1)}, \,
\kappa b \simeq \frac{3}{2} \sqrt{\dfrac{ e}{2 \rho}}
\end{equation}

\section{Numerical results and discussion}\label{sec:results}

 It is of considerable interest to compare exact numerical data with our analytical theory and to determine the regimes of validity of these asymptotic results. Here we present results of  numerical solutions of the system of Eqs.\eqref{Eq:PB_io} and \eqref{eq:Stokes} with prescribed boundary conditions, following the approach based on the collocation method~\cite{bader.g:1987}, together with specific calculations using asymptotic approximations.

\subsection{Velocity profiles}

\begin{figure}[t]
	\begin{center}
		\includegraphics[width=1\columnwidth]{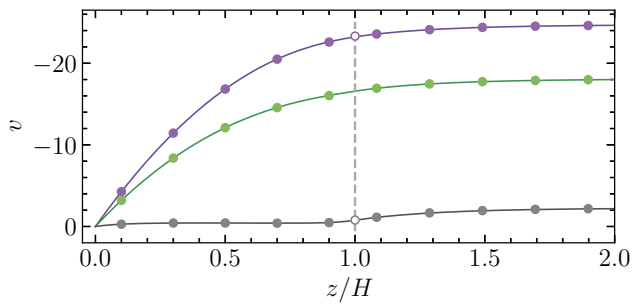}
	\end{center}
	\vspace{-0.4cm}
	\caption{The profiles $v$ computed using $\kappa H = 3$ and $\rho = 5$ with $\mathcal{K}H=0.1, 1,$ and 10 (solid curves from top to bottom). Filled circles show predictions of Eqs.~\eqref{eq:v_o},\eqref{eq:EO_in_thick}. Open circles correspond to $v_s$ calculated from Eqs.\eqref{eq:v_s_large} and \eqref{eq:v_s_thick_betagg1}. } \label{fig:Fig2}
\end{figure}

We begin by studying the velocity profiles at fixed $\rho = 5$, $\kappa H = 3$ and several $\mathcal{K} H$ in the range from 0.1 to 10. Numerical results are shown in Fig.~\ref{fig:Fig2}. Also included are theoretical curves calculated from Eqs.\eqref{eq:EO_in_thick} for $v_i$ and \eqref{eq:v_o} for $v_o$
(with $v_{\infty}$ defined by Eq.\eqref{eq:v_inf}). In the later case we used Eq.\eqref{eq:v_s_thick} to calculate $v_s$. These exact theoretical results fully coincide with numerical data. The calculations of $v_s$ in the limits of small and large $\mathcal{K}H$ are also included in Fig.~\ref{fig:Fig2} showing their coincidence with numerical solutions using $\mathcal{K} H = 0.1$ and 10. It can be seen that on reducing $\mathcal{K} H$ the value of $-v$ increases. All outer velocity profiles are of precisely the same shape, set by $\psi_o$ (see Eq.\eqref{eq:v_o}), so that the dramatic increase in $-v_{\infty}$ upon decreasing $\mathcal{K} H$ is induced by changes in $v_s$ only. The latter are associated with the inner flows discussed in Sec.~\ref{Surface_slip}.

% to test our analytical results, asymptotic analysis
\begin{figure}[t]
	\begin{center}
		\includegraphics[width=1\columnwidth]{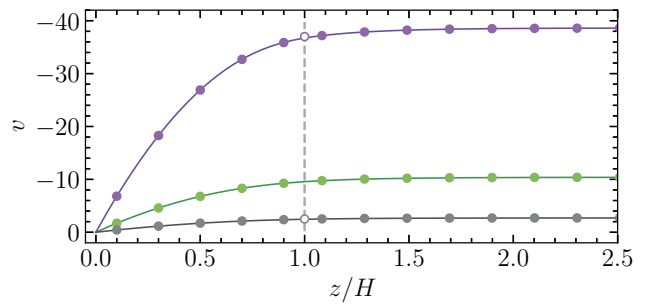}
	\end{center}
	\caption{The $v$-profiles computed using $\kappa H = 3$ and $\rho =8, 2, 0.5$ (solid curves from top to bottom), and $\mathcal{K}H = 0.1$. Filled circles show predictions of Eqs.~\eqref{eq:v_o} and \eqref{eq:EO_in_large_thick}. Open circles indicate $v_{s}$ calculated from Eqs.~\eqref{eq:v_s_large_smallrho} and \eqref{eq:v_s_large_largerho}.} \label{fig:Fig3a}
\end{figure}

To examine a significance of inner flows more
closely, in Fig.~\ref{fig:Fig3a} we plot the velocity profiles computed using $\mathcal{K}H = 0.1$. These results refer to fixed $\kappa H$, taken the same as in Fig.~\ref{fig:Fig2}, but we now vary $\rho$
from 0.5 to 8.  As $z/H$ is increased,  $-v$ grows strictly monotonically for all $\rho$, mostly in the inner (film) region, until it saturates in the bulk electrolyte. The parabolic shape of the inner profile is well seen in numerical examples with $\rho \geq 2$.
The magnitude of the velocity also grows on increasing $\rho$ confirming predictions of Sec.\ref{LBL}. Finally, we mention that calculations
made from Eqs.\eqref{eq:EO_in_large_thick} for $v_i$ and \eqref{eq:v_o} for $v_o$ fit perfectly the numerical data, so is $v_s$ obtained from Eqs.~\eqref{eq:v_s_large_smallrho} and \eqref{eq:v_s_large_largerho}.

\begin{figure}[t]
	\begin{center}
\includegraphics[width=1\columnwidth]{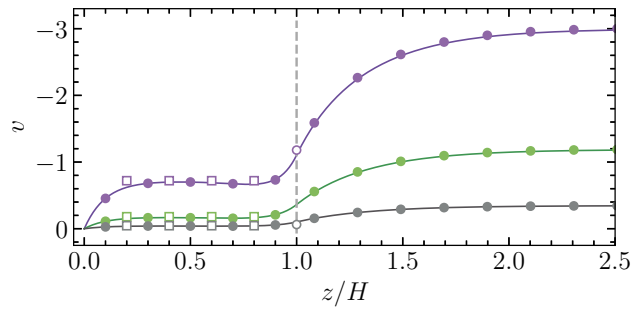}
	\end{center}
	\caption{The same as in Fig.~\ref{fig:Fig3a}, but for $\mathcal{K}H =10$. Filled circles show predictions of Eqs.~\eqref{eq:v_o} and \eqref{eq:SBL_vi}. Open circles show $v_{s}$ calculated from Eqs.~\eqref{eq:v_s_large_beta_rholl1} and \eqref{eq:v_s_large_beta_rhogg1}. Squares indicate $v_i$ from \eqref{eq:plateau0}.} \label{fig:Fig3b}
\end{figure}

The results of the identical computations, but made for large $\mathcal{K}H$, together with theoretical $v$, from Eqs.~\eqref{eq:v_o} and \eqref{eq:SBL_vi} are included in Fig.~\ref{fig:Fig3b}.  The absolute value of velocity shows a weakly monotonic grows with $z/H$, and saturates at infinity, i.e. outside of the outer diffuse layer. The inner velocity shows a distinct plateau, where
constant $v_i$ is given by Eq.\eqref{eq:plateau0}. On increasing $z/H$ further $-v_i$ grows up to $-v_{s}$, and the velocity jump in the interface layer is equal to $-\dfrac{\kappa _{i} }{\kappa _{i}+\mathcal{K}} \Delta \psi$ as follows from Eq.\eqref{eq:v_s_thick_betagg1}. Then in the outer region $-v_o$ increases from $-v_s$ until $-v_{\infty}$ in the electroneutral bulk. It can be seen that even at large $\mathcal{K}H$ slip velocity makes a discernible contribution to the flow enhancement, provided $\rho$ is large enough. This is exactly what we have proposed in Sec.~\ref{SBL}. Note, however, that the absolute values of velocities are an order of magnitude smaller than in
Fig.~\ref{fig:Fig3a}.

\subsection{Zeta-potential versus slip velocity}\label{sec:discussion_zeta}

We now turn to zeta-potential of surfaces. The main issue we address is how to enhance $\zeta$ by generating a large slip velocity at the surface.

\begin{figure}[h]
\begin{center}
\includegraphics[width=1\columnwidth]{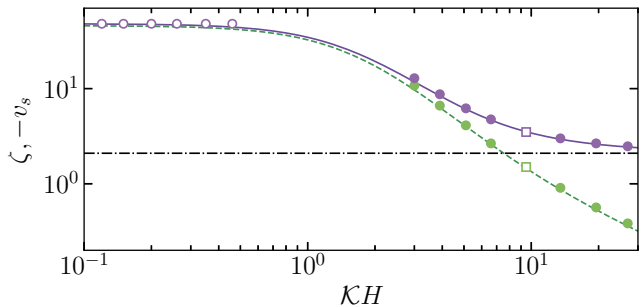}
\end{center}
\caption{Zeta-potential $\zeta$ (solid curve) and slip velocity $-v_s$ (dashed curve) computed for a film of $\kappa H = 3$ and $\rho = 10$ as a function of  $\mathcal{K}H$. The surface potential $\psi_{s}$ is shown by dash-dotted line. Open  circles indicate calculations from Eq.~\eqref{eq:zeta2}. Filled circles are obtained using Eqs.\eqref{eq:v_s_large_beta_rhogg1} and \eqref{eq:v_bulk_large_beta_rhogg1}. Squares show predictions of Eq.\eqref{eq:v_ki_rhogg1}. }
\label{fig:Fig4}
\end{figure}

Fig.~\ref{fig:Fig4}, plotted in a log-log scale, is intended to indicate the ranges of $-v_s$ and $\zeta = -v_{\infty}$ that are encountered at different $\mathcal{K} H$. For this numerical example we use a film of $\kappa H = 3$, as before, and explore only the case of a large $\rho = 10$, where the flow enhancement is more pronounced. The surface potential, $\psi_s$, of such a film is also shown in Fig.~\ref{fig:Fig4} and it is seen that it is quite small. $\zeta$ takes its maximal (constant, i.e. independent on the Brinkman length) values at $\mathcal{K} H \leq 1$. This part of the curve is well described by Eq.\eqref{eq:zeta2}, pointing out that this asymptotic approximation has validity well beyond the range of the original assumptions (see Sec.\ref{kineqK}). We remark and stress that for parameters, chosen for this specimen example, the upper value of $\zeta$ is several tens of times larger than $\psi_s$. Zeta-potential then reduces and meets $\psi_s$ tangentially at very large $\mathcal{K} H$. When $\mathcal{K} H \geq 2$, this decay is well consistent with predictions of Eq.\eqref{eq:v_bulk_large_beta_rhogg1}, thus the latter is also valid well outside the range of its formal applicability. One important conclusion from Fig.~\ref{fig:Fig4} is that $\zeta \simeq 10$ at $\mathcal{K} H = 2$ and can be several times larger that $\psi_s$ even when $\mathcal{K} H = O(10)$.
Another conclusion we would like to highlight is that when $\mathcal{K} H \leq 1$, $\zeta \simeq - v_s$, but when $\mathcal{K} H$ is in the range from ca. 4 to 20, zeta-potential is dictated both by $\psi_s$ and $v_s$.
 Finally, we recall that these asymptotic approximations have been derived assuming $ \kappa_i \neq \mathcal{K}$. The (smooth) numerical curves shown in Fig.~\ref{fig:Fig4} include $\zeta$ and $-v_s$ defined at $ \kappa_i = \mathcal{K}$, so that we have also calculated their values from Eq.\eqref{eq:v_ki_rhogg1} and see that they well agree with numerical data.

\begin{figure}[h]
\begin{center}
\includegraphics[width=1\columnwidth]{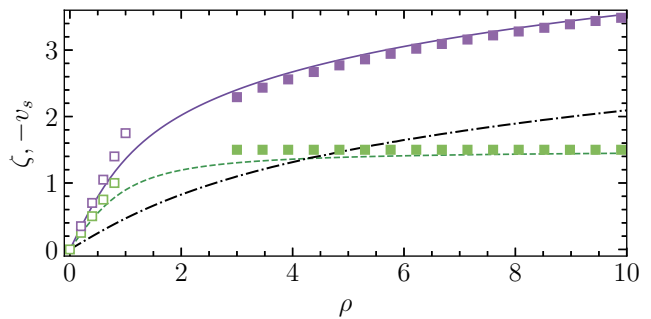}
\end{center}
\caption{Zeta-potential $\zeta$ (solid curve) and slip velocity $-v_s$ (dashed curve) vs. $\rho$, computed for the case of $\mathcal{K}=\kappa_{i}$ using $\kappa H = 3$. The surface potential $\psi_{s}$ is shown by dash-dotted line. Open squares show calculations from Eq.\eqref{eq:v_ki_rholl1}, filled squares are obtained from Eq.\eqref{eq:v_ki_rhogg1}.} \label{fig:Fig9}
\end{figure}

In the special case of $ \kappa_i = \mathcal{K}$ the zeta potential and slip velocity depend on $\rho$ only (see Sec.\ref{kieqK}). If we keep $\kappa H = 3$ fixed and increase $\rho$ from 0 to 10 imposing the condition $ \kappa_i = \mathcal{K}$, the value of $\mathcal{K} H$ will increase with $\rho$ from 3 to ca. 9.5. We thus obtain the situation illustrated by  Fig.\ref{fig:Fig9}. As $\rho$ is increased, both $\zeta$ and $\psi_s$ increase, but note that $\zeta$ always exceeds $\psi_s$. One can see, that at very small $\rho$ the contribution of slip velocity to $\zeta$ is significant, so that it becomes several times larger than $\psi_s$. This part of the curves is well described by Eqs.\eqref{eq:v_ki_rholl1}. Already when $\rho \simeq 3$, $-v_s$ saturates, but $\zeta$ grows further, solely since $\psi_s$ continues to increase with $\rho$. Asymptotic approximations, Eqs.\eqref{eq:v_ki_rhogg1}, although obtained for $\rho \gg 1$, fit our numerical results well, when $\rho \geq 3$.

\subsection{Amplification factor}

\begin{figure}[t]
\begin{center}
\includegraphics[width=1\columnwidth]{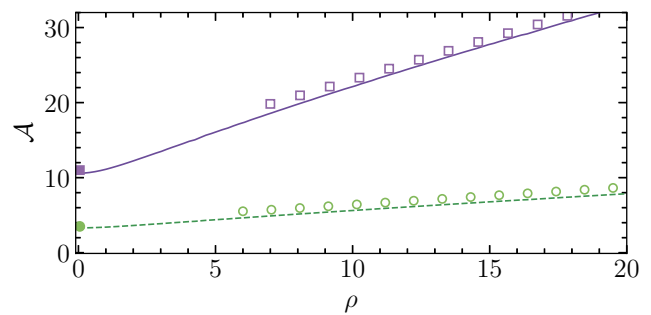}
\end{center}
\caption{The amplification factor $\mathcal{A}$  computed as a function of $\rho$ for a film of $\kappa H = 3$ using $\mathcal{K} H = 0.3$ (solid curve) and $\mathcal{K}H = 3$ (dashed curve).  Open and filled squares show predictions of Eqs.~\eqref{eq:zeta_large_rholl1} and \eqref{eq:zeta_large_rhogg1}. Open and filled circles show results of calculations from Eqs.~\eqref{eq:A_small_rholl1} and \eqref{eq:A_large_beta_rhogg1}.}
 \label{fig:Fig5}
\end{figure}

Next we examine the amplification factor that characterizes an enhancement of an outer plug flow relative to what is generated near surfaces of  $\zeta \simeq \psi_s$. Since $\mathcal{A}$ is defined by Eq.\eqref{eq:amplification} one can immediately calculate it using the results of Sec.\ref{sec:discussion_zeta}. Our theory suggests several distinct routes  to increase $\mathcal{A}$, which implies that there is a choice of appropriate variables for illustrating them.

The above results (see Fig.~\ref{fig:Fig4}) indicate that our asymptotic  equations could be applicable for calculating $\mathcal{A}$ at certain   values of $\mathcal{K}H$, intermediate
between those appropriate to Figs.~\ref{fig:Fig3a} and \ref{fig:Fig3b}. We now set $\mathcal{K}H = 0.3$ and 3 that may be considered as intermediate, and  in Fig.\ref{fig:Fig5} show  $\mathcal{A}$ computed using $\rho$ as a variable. It is interesting to note that the amplification factor is finite at $\rho = 0$, when $\psi_s$ vanishes. It increases with $\rho$ for both $\mathcal{K}H$, and when $\rho \geq 5$, the growth appears as linear. The flow is amplified in both cases, but much stronger when $\mathcal{K}H = 0.3$ leading to $\mathcal{A} = O(10)$. Overall, the agreement with asymptotic expressions obtained for small and large $\mathcal{K}H$ is very good.

\begin{figure}[h]
\begin{center}
\includegraphics[width=1\columnwidth]{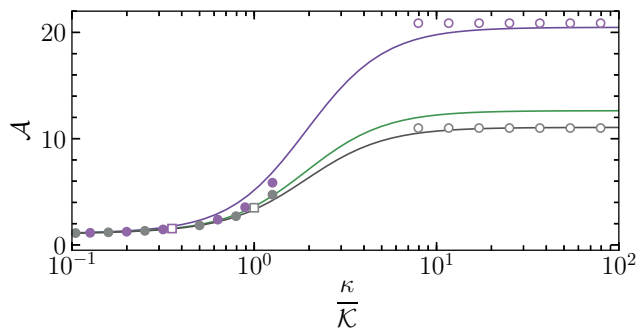}
\end{center}
\caption{The amplification factor $\mathcal{A}$  computed as a function of $\kappa/\mathcal{K}$ for a film of $\kappa H = 3$ using $\rho = 8,$ $2,$ and $0.5$ (solid curves from top to bottom). Filled circles show calculations from Eqs.\eqref{eq:A_small_rholl1} and \eqref{eq:A_large_beta_rhogg1}, open circles - from Eqs.\eqref{eq:zeta_large_rholl1} and \eqref{eq:zeta_large_rhogg1}.  Squares indicate $\mathcal{A}$ from Eqs.\eqref{eq:A_ki_rholl1} and \eqref{eq:A_ki_rhogg1}.} \label{fig:Fig7}
\end{figure}

Another parameter that controls $\mathcal{A}$ is $\varkappa$ given by Eq.\eqref{eq:varkappa}, which is the ratio of the Brinkman length to the Debye length. When it is small, only a marginal flow amplification could be expected, and this immediately interprets the lower curve in Fig.\ref{fig:Fig5}, where $\varkappa = 1$. However, when this ratio is large, the flow should be significantly amplified. This is exactly what we observe for the upper curve in Fig.~\ref{fig:Fig5} corresponding to $\varkappa = 10$ and relatively small $\mathcal{K} H$. As follows from Eq.\eqref{eq:A_small_rholl1}, in the case of $\varkappa \gg 1$ a very large $\mathcal{A}$ can be expected even when the Brinkman screening length and $\rho$ are relatively small. This is illustrated in Fig.~\ref{fig:Fig7}, where we plot $\mathcal{A}$ as a function of $\varkappa$ for several $\rho$. In a lin-log scale of Fig.~\ref{fig:Fig7} the curves appear (shifted and scaled) sigmoid, i.e. having a characteristic inclined `S'-shape with a bell shaped first derivative  (not shown), and are constrained between $\mathcal{A} = 1$ and its maximal, insensitive to further increase in $\varkappa$,  value. The latter increases with $\rho$ and we see that the data for $\rho = 0.5$ and $8$ are well fitted by Eqs.\eqref{eq:zeta_large_rholl1} and \eqref{eq:zeta_large_rhogg1}, respectively. As predicted in Sec.\ref{SBL}, at relatively small $\rho$ we see an order of magnitude flow amplification when $\varkappa \geq 8$.
The lowermost branches of these two curves are in agreement with Eqs.\eqref{eq:A_small_rholl1} and \eqref{eq:A_large_beta_rhogg1}, and Eqs.\eqref{eq:A_ki_rholl1} and \eqref{eq:A_ki_rhogg1} are equally valid for the case of $\kappa_i \neq \mathcal{K}$. Finally we note that $\mathcal{A}$ is pretty large already when $\varkappa = O(1)$.

\begin{figure}[t]
\begin{center}
\includegraphics[width=1\columnwidth]{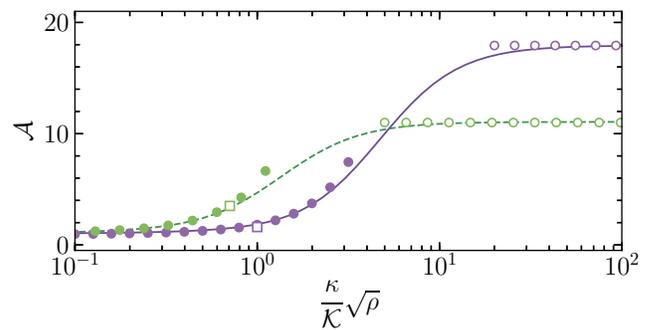}
\end{center}
\caption{The amplification factor $\mathcal{A}$ as a function of $\sqrt{\rho}\varkappa$ computed for $\kappa H = 3$ and $\rho = 0.5$ (dashed line), $\rho = 6$ (solid line). Open circles show calculations from Eqs.\eqref{eq:zeta_large_rholl1} and \eqref{eq:zeta_large_rhogg1}. Filled circles are obtained from Eqs.\eqref{eq:A_small_rholl1} and \eqref{eq:A_large_beta_rhogg1}. Squares indicate calculations from Eqs.~\eqref{eq:A_ki_rholl1} and \eqref{eq:A_ki_rhogg1}.} \label{fig:Fig6}
\end{figure}

The volume charge density $\rho$ and $\varkappa$ are parameters that control  $\mathcal{A}$. Eq.\eqref{eq:A_large_beta_rhogg1} indicates that their combination, $\varkappa \sqrt{\rho}$, could be an equally useful variable and provides an additional insight to the problem. In Fig.~\ref{fig:Fig6} we plot $\mathcal{A}$ versus $\varkappa \sqrt{\rho}$. The numerical calculations are made using quite small and moderate $\rho$, and we conclude that both curves are again sigmoidal. Note that the inflection point is located at smaller $\varkappa \sqrt{\rho}$ if $\rho = 0.5$. Another unexpected result is that the amplification factor for a weakly charged film is larger than that for a coating of $\rho = 6$ until the two curves meet at certain value of $\varkappa \sqrt{\rho}$ ($\simeq 6$ with these parameters). Only on increasing $\varkappa \sqrt{\rho}$ further a stronger charged film provides a better amplification of the flow. The agreement with asymptotic approximations is again very good and we omit a detailed discussion of that.

\subsection{Electro-osmotic slip length}

In Sec.\ref{sec:general}, we suggested an equivalent interpretation of flow enhancements in terms of an electro-osmotic slip length, which is established self-consistently. In contrast to  $\mathcal{A}$, which characterizes the flow enhancement in the bulk, the slip length is the property of the interface itself and is related to $\mathcal{A}$ by Eq.\eqref{eq:hydrophobic}. Since the main reason for a flow enhancement is the ratio of $b$ to Debye length, below we investigate the variation of $\kappa b$ in response to changes in $\rho$ and $\mathcal{K}H$.

\begin{figure}[h]
\begin{center}
\includegraphics[width=1\columnwidth]{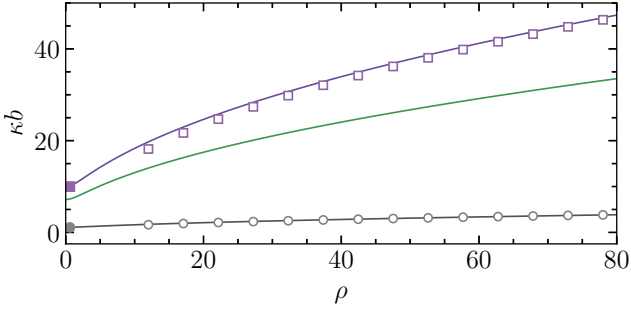}
\end{center}
\caption{Scaled slip length $\kappa b$ computed for a film of $\kappa H = 3$ as a function of $\rho$ using $\mathcal{K}H = 0.1$, $1$, and $10$ (solid curves from top to bottom). Open and filled squares show calculations from Eqs.\eqref{eq:b_large_rholl1} and \eqref{eq:b_large_rhogg1}. Open and filled circles indicate results of Eqs.\eqref{eq:b_small_rholl1} and \eqref{eq:b_small_rhogg1}.}
\label{fig:Fig_slip1}
\end{figure}

Numerical results for $\kappa b$ as a function of $\rho$, obtained
with different $\mathcal{K}H$ are shown in Fig.\ref{fig:Fig_slip1}.
The asymptotic approximations are very well verified for large and small values of $\mathcal{K}H$, both for a limit of vanishing $\rho$ (Eqs.\eqref{eq:b_large_rholl1} and \eqref{eq:b_large_rhogg1}) and for highly charged films (Eqs.\eqref{eq:b_small_rholl1} and \eqref{eq:b_small_rhogg1}). It is well seen in Fig.\ref{fig:Fig_slip1} that $\kappa b$ of porous surfaces  grows  with a volume charge density, and we remark that at low $\mathcal{K}H$ it increases more rapidly with increasing $\rho$, especially for weakly charged films.

\begin{figure}[h]
\begin{center}
\includegraphics[width=1\columnwidth]{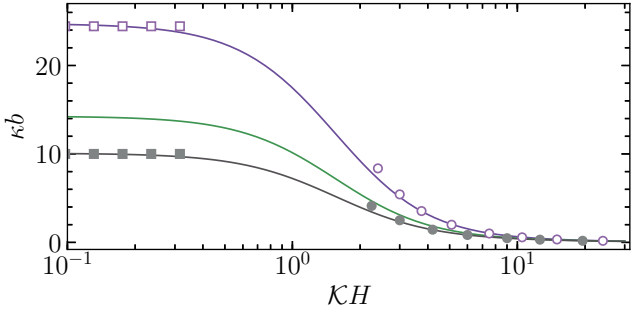}
\end{center}
\caption{Scaled slip length $\kappa b$ computed for a film of $\kappa H = 3$ as a function of $\mathcal{K}H$ using $\rho = 10,$ $5,$ and $0.5$ (solid curves from top to bottom). Open and filled squares show calculations from Eqs.\eqref{eq:b_large_rholl1} and \eqref{eq:b_large_rhogg1}. Open and filled circles indicate results obtained from Eqs.\eqref{eq:b_small_rholl1} and \eqref{eq:b_small_rhogg1}.}
\label{fig:Fig_slip2}
\end{figure}

By varying $\mathcal{K}H$ at fixed $\rho$ it is possible to obtain the curves for $\kappa b$ shown in Fig.\ref{fig:Fig_slip2}. The curves plotted in a lin-log scale resemble (shifted and scaled) inverse sigmoids smoothly decaying from low down to large $\mathcal{K} H$ regimes on increasing  $\mathcal{K} H$ from ca. 0.3 to 2. Outside of this transient range of $\mathcal{K} H$ the asymptotic approximations appear to adequately describe numerical data.

\section{Towards tuning  zeta-potential  by salt.}\label{sec:salt}

 So far we have considered the electro-osmotic properties in terms of dimensionless generic parameters, such as $\rho$, $\kappa H$, $\mathcal{K}H$, and their combinations. Additional insight into the problem can be gleaned by calculating $\zeta$ as a function of $c_{\infty} \propto \kappa^{2}$ at given $H$, $\Lambda$, and $\varrho$.

 Let us now keep fixed $H = 100$ nm and set $\Lambda = 200$ and 20 nm, which correspond to $\mathcal{K}H = 0.5$ and 5. We also keep fixed $\varrho = 150$ kC/m$^3$, which is close to reported in experiment by~\citet{duval.j:2009}, and
  vary $c_{\infty}$ from $10^{-5}$ to $10^{-2}$ mol/L.
Upon increasing $c_{\infty}$ in this range,  $\rho$ is reduced from about 78 down to 0.08,
so that regimes of highly and weakly charged coatings can be  tuned simply by adjusting the concentration of salt, and $\kappa_i H$ is increased from about
3 to 33. Thus, the criterion of a thick film, Eq.\eqref{eq:criterion}, is  fulfilled strictly in our examples, except concentrations that are very close to  $10^{-5}$ mol/L.

\begin{figure}[h]
	\begin{center}
		\includegraphics[width=1\columnwidth]{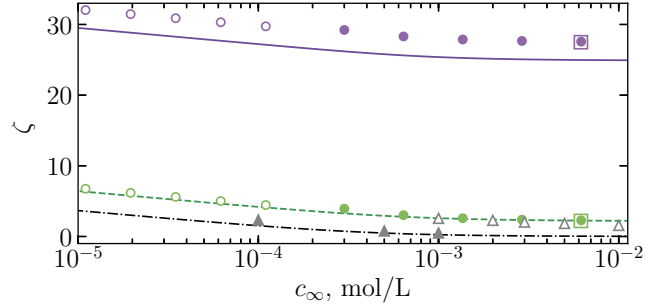}
	\end{center}
	\caption{Zeta-potential $\zeta$ vs. $c_{\infty}$, computed using $H = 100$ nm, $\varrho = 150$ kC/m$^3$ with $\Lambda = 20$ nm (solid curve) and $\Lambda = 200$ nm (dashed curve). The surface potential $\psi_{s}$ is shown by dash-dotted line. Open circles are obtained from Eqs.\eqref{eq:log_plateau} and \eqref{eq:log_plateau_large}. Filled circles show calculations from Eqs.\eqref{eq:plateau} and \eqref{eq:plateau_large}. Large squares correspond to calculations using second term of Eqs.\eqref{eq:plateau} and \eqref{eq:plateau_large} only. Filled and open triangles indicate experimental data~\cite{sobolev.vd:2017,lorenzetti.m:2016}.}
		\label{fig:fig_zeta_salt}
\end{figure}

The computed $\zeta$ is shown in Fig.~\ref{fig:fig_zeta_salt} together with $\psi_s$. We recall that $\psi_s$ can be considered as $\zeta$ of a reference surface of $\Lambda = 0$. The surface potential  reduces from 4 ($\Psi_s \simeq 100$ mV) practically to zero as $c_{\infty}$ increases, leading to a suppression of a flow. In dilute solutions, where the film is highly charged, the decay is logarithmic, $\psi_s \simeq \ln{\left(\dfrac{\varrho}{\textsl{e} c_{\infty}}\right)}-1$, and appears as linear in this lin-log plot~\cite{silkina.ef:2020b}. Since the coating becomes weakly charged in more concentrated solutions, $\psi_s \simeq \dfrac{\varrho}{4 \textsl{e} c_{\infty}}$, as follows from Eq.\eqref{eq:small_rho_summary}.
When $\Lambda$ is finite, zeta-potential becomes larger, and the curves computed using $\Lambda = 20$ and 200 nm look like the  shifted $\psi_s$-curve.

In the situation of $\Lambda = 200$ nm and low salt we can use Eq.\eqref{eq:zeta2} to obtain
\begin{equation}\label{eq:log_plateau}
    \zeta \simeq \ln{\left(\frac{\varrho}{\textsl{e} c_{\infty}}\right)}  + \dfrac{2\pi\ell_{B}\varrho H^2}{\textsl{e}}
\end{equation}
This equation is equivalent to derived by \citet{silkina.ef:2020b} for the case of $\Lambda \to \infty$.
 From Eq.\eqref{eq:zeta1},
 \begin{equation}\label{eq:plateau}
    \zeta \simeq \frac{\varrho}{2 \textsl{e} c_{\infty}}  + \dfrac{ 2 \pi \ell_B   \varrho H^2}{\textsl{e}},
\end{equation}
which describes a salt dependence of $\zeta$ in concentrated solutions and large Brinkman lengths.
For sufficiently large $c_{\infty}$ the first term is negligibly small, and $\zeta$ becomes constant. The nature of this constant, i.e. of the second term in Eqs.\eqref{eq:log_plateau} and \eqref{eq:plateau}, is apparent. Set $\Delta \psi = 0$, Eq.\eqref{eq:v_s_large} then yields a value for $-v_s$, and we conclude that the later depends only on the volume charge density and hydrodynamic permeability.
Of course, Eqs.\eqref{eq:log_plateau} and \eqref{eq:plateau} are very approximate, when $\Lambda$ is not too large compared to $H$,
but it provides us with some guidance.
Indeed, it is seen in Fig.~\ref{fig:fig_zeta_salt} that they overestimate numerical data for $\Lambda = 200$ nm. A more precise description requires  accounting of a higher-order term into expansion \eqref{eq:EO_in_large_thick} which is beyond the scope of the present work.

When $\Lambda = 20$ nm, for dilute solutions we can use Eq.\eqref{eq:v_bulk_large_beta_rhogg1}. An order of magnitudes estimate shows that with our parameters $\varkappa \sqrt{\rho} \simeq 1.8$, so that the contribution of the first term is practically compensated by that of the last one, so that
\begin{equation}\label{eq:log_plateau_large}
  \zeta \simeq \ln{\left(\frac{\varrho}{\textsl{e} c_{\infty}}\right)}  + \dfrac{4 \pi \ell_{B} \Lambda^2 \varrho}{\textsl{e}}
\end{equation}
For concentrated solutions   Eq.\eqref{eq:zeta3} should be employed, and, using that at high salt $\varkappa$ becomes large, we find
\begin{equation}\label{eq:plateau_large}
  \zeta \simeq \frac{\varrho}{2 \textsl{e} c_{\infty}}  + \dfrac{4 \pi \ell_{B} \Lambda^2 \varrho}{\textsl{e}}
\end{equation}
Fig.~\ref{fig:fig_zeta_salt} shows that calculations from Eqs.\eqref{eq:log_plateau_large} and \eqref{eq:plateau_large} are in excellent agreement with numerical results for $\Lambda = 20$ nm. The second term in Eqs.\eqref{eq:log_plateau_large} and \eqref{eq:plateau_large} represents $- v_s$ given by Eq.\eqref{eq:v_s_large_beta_rholl1} if and only if $\Delta \psi = 0$, i.e. at high salt concentration. We again conclude that it is defined by the volume charge density and hydrodynamic (now Darcy) permeability, and note that in such a situation the large zeta-potential reflects solely an  electro-osmotic plug flow inside the film (the latter is well seen in Fig.~\ref{fig:Fig3b}).

There have been experimental reports of a phenomenon of a finite electrokinetic mobility in concentrated solutions for various systems, including the cell surface coated with
a layer of charged glycoproteins and glycolipids~\cite{donath.e:1979}, `hairy' polystyrene latexes~\cite{garg.a:2016}, and more~\cite{irigoyen.j:2009,irigoyen.j:2013}, but it has never been interpreted in this fashion. \citet{sobolev.vd:2017} recently reported zeta-potential measurements for a set of polyacrylonitrile hollow fiber  membranes by varying $c_{\infty}$ from $10^{-4}$ to $10^{-3}$ mol/L and concluded that $\zeta$ reduces with salt. Their experimental results, included in Fig.~\ref{fig:fig_zeta_salt}, are similar to the portion of our numerical curve corresponding to their concentration range, but the values of $\zeta$ are below our calculations due to smaller experimental $\psi_s$. Note that their experimental values of $\mathcal{A}$ grow from about 1.5 to 2 and well reproduce the trend we predicted, but are of smaller values than calculated with our parameters theoretically (not shown).
\citet{lorenzetti.m:2016} have obtained $\zeta$ vs. $c_{\infty}$ curves for nanotubular surfaces, proposed as coating material for Ti body implants, by varying $c_{\infty}$ from $10^{-3}$ to $10^{-2}$ mol/L. As a side note,  the authors~\cite {lorenzetti.m:2016} did not attempt to make connection with the volume charge density and $\Lambda$, so the status of their theory remains obscure. Their experimental results are, however, consistent with Eq.\eqref{eq:plateau_large} and practically coincide with our numerical curve.

\begin{figure}[h]
	\begin{center}
		\includegraphics[width=1\columnwidth]{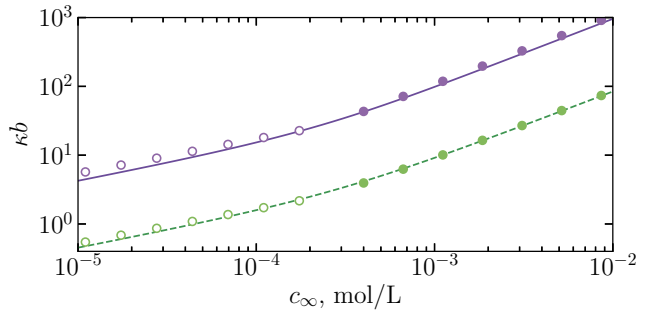}
	\end{center}
	\caption{Scaled slip length $\kappa b$ computed as a function of $c_{\infty}$ for a film of $H = 100$ nm, $\varrho  = 150$ kC/m$^3$ using $\Lambda = 200$ nm (solid curve) and 20 nm (dashed curve). Open circles show calculations from Eqs.\eqref{eq:kb_low_salt}. Filled circles are obtained from Eqs.\eqref{eq:kb_high_salt}. }
	\label{fig:fig_kb_salt}
\end{figure}

Such a behavior can be interpreted in terms of the slip length, which is related to $\mathcal{A}$ by Eq.\eqref{eq:hydrophobic}. Fig.~\ref{fig:fig_kb_salt} shows that $\kappa b$ grows with salt, and  there are again two distinct regimes. At low concentrations
\begin{equation}\label{eq:kb_low_salt}
  \kappa b \simeq 2 \pi \ell_{B} H^2 \sqrt{\dfrac{e \varrho c_{\infty}}{\textsl{e}}}, \, \kappa b \simeq 4 \pi \ell_{B} \Lambda^2 \sqrt{\dfrac{e \varrho c_{\infty}}{\textsl{e}}}
\end{equation}
 Here the first equation corresponds to large, and the second to small $\Lambda$. We see that $\kappa b$ grows linearly with the hydrodynamic permeability of the coating and that $\kappa b \propto \sqrt{\varrho c_{\infty}/{\textsl{e}}}$ (or $ b \propto \sqrt{\varrho /{\textsl{e}}}$).
At high concentrations
\begin{equation}\label{eq:kb_high_salt}
\kappa b \simeq 8 \pi \ell_{B} H^2 c_{\infty}, \, \kappa b \simeq 16 \pi \ell_{B} \Lambda^2 c_{\infty},
\end{equation}
i.e. in a high salt regime $\kappa b \propto c_{\infty}$ (or $ b \propto \sqrt{c_{\infty}}$) and does not depend on the volume charge density of the film. The linear increase with the hydrodynamic permeability remains. Note that in this situation $\zeta \simeq -v_s \simeq - \kappa b \varrho /{(4\textsl{e} c_{\infty})}$ and $\kappa b \simeq \mathcal{A}$. Thus, at high salt both $\kappa b$ and $\mathcal{A}$ are actually very large, up to $10^2-10^3$ as seen in Fig.~\ref{fig:fig_kb_salt}.

\section{Concluding remarks}\label{sec:conclusion}

Our model, which is probably the simplest realistic model for the electro-osmosis near supported porous films that one might contemplate, provides considerable insight into different regimes of the flow and suggests several routes for its enhancement. Many of our results will have validity beyond some specific assumptions underlying the analysis, as demonstrated by numerical calculations.

The main results of our work can be summarized as follows. The Smoluchowski theory of electro-osmosis cannot
be employed for the description of zeta-potential of porous surface, except some specific cases in the situation when $\Lambda \ll H$. Namely, to provide $\zeta \simeq \psi_s$ for weakly charged films the Brinkman length $\Lambda$ should be much smaller than $\lambda_D$, but for highly charged coatings, the condition $ 2 \Lambda \sqrt{\dfrac{\pi \ell_{B} \varrho}{\textsl{e}} }\ll 1$ should be fulfilled.
In all other cases $\zeta$ may be significantly augmented compared to the surface potential due to liquid slip at the
porous surface. Even when $\Lambda$ is small, $\zeta$ can become very large provided both $\dfrac{\varrho}{2 \textsl{e} c_{\infty}} $ and $ 2 \Lambda \sqrt{\dfrac{\pi \ell_{B} \varrho}{\textsl{e}} } \gg 1$.
The slip velocity emerges due to electro-osmotic flow inside the porous film (that is associated with an electric body
force on the accumulated in the Donnan area mobile counter-ions), and  depending on the value of $\Lambda$, two different scenarios occur. For large $\Lambda$, an inner flow resembles the usual parabolic
Poiseuille flow leading to a very large slip velocity, but for small $\Lambda$ it is similar to the Darcy plug velocity in a pressure-driven flow.  Thus, for porous media the origin of electro-osmotic flows should be attributed both to diffuse layers, and the absorbed ions. The former mechanism is, of course, traditional, while the latter is specific for porous surfaces only. It is responsible for a flow amplification and dominates at high salt.

We have proposed at the beginning that it may be convenient to quantify the slip velocity at the surface in terms of the slip length. Such an approach has been successfully employed  in describing flows in various polymer systems~\cite{brochard.f:1992,horn.rg:2000,barraud.c:2019}, and those over hydrophobic surfaces, where the slip length $b$ reflects the wettability, but does not depend on the salt concentration~\cite{audry.mc:2010,joly.l:2006b}. \citet{joly.l:2006b} also suggested that $b$ of impermeable surfaces could slightly decrease with the surface charge density. By contrast, the  electro-osmotic  slip length of permeable surfaces  is proportional to  the square root of $\varrho$ at low salt, and grows with the square root of $c_{\infty}$ in rather concentrated electrolyte solutions.

As mentioned in Sec.~\ref{sec:introduction}, for microfluidic applications it would be important to achieve velocities of a few
millimeters per second at a low-voltage. Returning to dimensional variables and using the results of Sec.~\ref{sec:salt} we conclude that when the Brinkman length is large, $V_s \simeq \dfrac{H^2 \varrho E}{2 \eta}$. Then using the dynamic viscosity of water and typical (low) $E = -3$ kV/m, we obtain $V_s \simeq -2.3$ mm/s for a film of $\Lambda = 200$ nm described in Sec.~\ref{sec:salt}. For the second film analyzed in the same section,  of $\Lambda = 20$ nm,
$V_s \simeq \dfrac{\Lambda^2 \varrho E}{\eta}$, and $V_s \simeq - 0.2$ mm/s. This velocity, of course, could  be enhanced for films of larger $\varrho$, but probably not much. The largest value of $\varrho$ we found in the literature was about 400 kC/m$^3$ (for polyacrylamide gels)~\cite{yezek.lp:2005}, which has potential to provide only twice faster slip velocity compared to our examples. To increase $V_s$ a more efficient strategy would be increasing the hydrodynamic permeability, and we will return to this point below.

Several of our theoretical predictions could be tested by experiment. We have already mentioned the work of \citet{donath.e:1979} who found a finite $V_{\infty}$ at high salt, as well as the papers by \citet{sobolev.vd:2017} and \citet{lorenzetti.m:2016} who reported the same trends as we predict here,  but they used a very narrow range of $c_{\infty}$.
The concentration must vary in the range described in Sec.~\ref{sec:salt} and Fig.~\ref{fig:fig_zeta_salt} to find a transition between highly and weakly charged film regimes. It would be of some interest to find $\varrho$ from the slope of the $\zeta$-curve at low salt concentrations, from Eq.\eqref{eq:log_plateau_large}. Once it is known, the value of $\Lambda$ can be obtained using Eq.\eqref{eq:log_plateau_large}, from the high salt plateau, as well as salt dependence of $\psi_s$ can be easily determined using expressions given in Sec.~\ref{sec:salt} (and then may be verified by using Eq.\eqref{eq:psi_s_thick}). It should be possible to obtain $\psi_s$ from electrostatic force measurements using the atomic force microscope~\cite{butt.hj:2005}, but this would invoke, beside tedious experiments, a very complicated data analysis with computational efforts~\cite{ohshima.h:2006}, so that electrokinetic measurements of $\psi_s$ appear easier and more promising.

The hydrodynamic permeability  depends on the porous texture (volume fraction of the solid, size of the pores, their geometry, etc). The experimental examples discussed in the present paper correspond to quite dense textures, where $\Lambda$ tends to a small value. However, the implications of porosity for the inner electro-osmosis, which is  especially
important for inducing a very large zeta-potential, so far remain unexplored. Thanks to techniques coming from microelectronics and 3D printing, one can fabricate porous films, structured in a very well controlled way (often at the
micro- and nanometer scale). It would be of some interest to measure zeta-potential of more dilute textures such as arrays of pillars or periodic honeycomb structures, where larger $\Lambda$ could be expected.   To guide the optimal design of porous coatings to increase $\Lambda$ and for a consequent electro-osmotic flow amplification, it would be timely to develop  the theoretical approaches, similar to known in superhydrophobic microfluidics~\cite{vinogradova.oi:2011}.

Our strategy can be extended  to describe the amplification of a  variety of electrokinetic phenomena~\cite{anderson.jl:1989}, including the classical streaming current and less widely known diffusio-osmosis, where flow is driven by the gradient of salt concentration that produces electric field. The former has already been studied for some porous systems, such as hydrogel films~\cite{duval.j:2009} and membranes~\cite{sobolev.vd:2017}. It would appear that the trends that are observed are consistent with our theoretical description of the zeta-potential.
 We suggest that further analysis of these measurements, or similar measurements with other porous films in the  specified above concentration range should employ our interpretation of zeta-potential, rather than the Smoluchowski approach since the latter makes a no-slip assumption, which is not generally valid for porous surfaces.
 The information about diffusio-osmotic flow near porous surfaces is still rather scarce.
 Although we are not aware of any direct measurements, some indirect experiments have recently revealed extremely strong and long-range diffusio-osmotic flows in the presence of porous surfaces, that have consequences for the remote control of particles assemblies~\cite{feldmann.d:2020} and even for  laundry cleaning~\cite{shin.s:2018}. Interestingly, the logarithmic decay fits well the diffusio-osmotic velocities obtained for real porous materials in dilute solutions by \citet{feldmann.d:2020}.
 The measurements of diffusio-phoresis of latex particles by \citet{prieve.d:2019} also lend some support to the picture of the slip velocity that is presented here. A systematic study of the diffusio-osmotic flow amplification by porous surfaces appears to be very timely and would constitute a significant extension of the present work.

\begin{acknowledgments}

This work was supported by the Ministry of Science and Higher Education of the Russian Federation and by the German Research Foundation (grant 243/4-2) within the Priority Programme ``Microswimmers - From Single Particle Motion to Collective Behaviour'' (SPP 1726).
\end{acknowledgments}

%\appendix

%\clearpage

%\bibliographystyle{apsrev4-1}
%%\bibliographystyle{unsrt}
%\bibliography{Striped_Gel}
%\bibliographystyle{plain}
%\bibliographystyle{plain}
\bibliography{Biblio_EO_Bag}

\end{document}